\documentclass[conference]{IEEEtran}

\newcommand{\nop}[1]{} 
\newcommand{\shorten}[1]{}
\usepackage{subfigure}
\usepackage{algorithmicx}
\usepackage{siunitx}
\usepackage{algpseudocode}
\usepackage[Algorithm,ruled]{algorithm}
\usepackage{tikz}
\usepackage{comment}
\usetikzlibrary{shapes,arrows}
\usepackage[T1]{fontenc}
\usepackage{subfig}
\usepackage{cite}      

\usepackage{graphicx}  

\usepackage{blindtext}
\usepackage{scrextend}
\addtokomafont{labelinglabel}{\sffamily}

\usepackage{psfrag}    

\usepackage{url}       
\usepackage{tikz}

\usepackage{xfrac}

\usepackage{amsmath}   
\usepackage{amssymb}

\newtheorem{example}{Example}

\newcommand{\signed}%
{{\unskip\nobreak\hfill\penalty50
		\hskip2em\hbox{}\nobreak\hfil $\blacksquare$
		\parfillskip=0pt \finalhyphendemerits=0 \par}}

\begin{document}

	\title{A Load Balancing Algorithm for Resource Allocation in IEEE 802.15.4e Networks}

	%
	\author{
		\IEEEauthorblockN{Katina Kralevska\IEEEauthorrefmark{1}, Dimitrios~J.~Vergados\IEEEauthorrefmark{2}, Yuming~Jiang\IEEEauthorrefmark{1}, and Angelos~Michalas\IEEEauthorrefmark{3}\\ Email: katinak@ntnu.no, djvergad@gmail.com, jiang@ntnu.no, amichalas@kastoria.teikoz.gr}
		\IEEEauthorblockA{\IEEEauthorrefmark{1}Department of Information Security and Communication Technology, 
			NTNU, Norwegian University of Science and Technology}
		\IEEEauthorblockA{\IEEEauthorrefmark{2}School of Electrical and Computer
			Engineering, National Technical University of Athens}
		\IEEEauthorblockA{\IEEEauthorrefmark{3}Department of Informatics Engineering, Technological Education Institute of Western Macedonia, Kastoria, Greece}
	}
	
	
	
	\maketitle

	\begin{abstract}
		The recently created IETF 6TiSCH working group combines the high reliability and low-energy consumption of IEEE 802.15.4e Time Slotted Channel Hopping with IPv6 for industrial Internet of Things. We propose a distributed link scheduling algorithm, called Local Voting, for 6TiSCH networks that adapts the schedule to the network conditions. The algorithm tries to equalize the link load (defined as the ratio of the queue length over the number of allocated cells) through cell reallocation. Local Voting calculates the number of cells to be added or released by the 6TiSCH Operation Sublayer (6top). 
		Compared to a representative algorithm from the literature, Local Voting provides simultaneously high reliability and low end-to-end latency while consuming significantly less energy. Its performance has been examined and compared to On-the-fly algorithm in 6TiSCH simulator by modeling an industrial environment with 50 sensors.  
	\end{abstract}

	
	%
	\IEEEpeerreviewmaketitle
	
	\section{Introduction}
	
	The Internet of Things (IoT) has boosted the deployment of Low-power and Lossy Networks (LLNs)\cite{ATZORI20102787}.
	LLNs consist of low complexity resource constrained embedded devices that are interconnected with different technologies.
	The IEEE 802.15.4e standard \cite{series/asc/GuglielmoAS14} defines the physical and the medium access (MAC) layers for LLNs.
	There are five MAC modes, including Time Slotted Channel Hopping (TSCH)\cite{rfc7554}. TSCH combines channel hopping and time synchronization where 
	all nodes in the network follow a common schedule that specifies for each node on which channel and at which time slot to communicate with its neighbors. IEEE 802.15.4e standard defines how the schedule is executed but it does not define how the schedule is built and updated.
	
	The IETF 6TiSCH working group \cite{wang-6tisch-6top-sublayer-04} defines mechanisms to combine the high reliability and low-energy consumption of IEEE 802.15.4e TSCH with the ease of interoperability and integration offered by the IP protocol. 
	6TiSCH Operation Sublayer (6top) integrates the IEEE 802.15.4e TSCH MAC layer with an IPv6-enabled upper stack. 
	6top includes a 6top Scheduling Function (SF) that defines the policy of adding/deleting TSCH cells between neighboring nodes while monitoring performance and collecting statistics. 
\shorten{		
	\begin{figure}
		\centering
		\includegraphics[width=0.31\textwidth]{protocolStack}
		\vspace{-0.1cm}
		\caption{The 6TiSCH protocol stack. We propose a scheduling algorithm that is located on the top of the 6top sublayer (presented in blue).}
		\vspace{-0.5cm}
		\label{layer}
	\end{figure}
}	
	On-the-fly (OTF) \cite{7273816} is a distributed algorithm for bandwidth allocation that calculates the number of cells to be added or deleted according to a neighbor-specific threshold. 
	OTF is prone to schedule collisions since nodes might not be aware of the cells allocated to other pairs of nodes. 
	Decentralized Broadcast-based Scheduling algorithm called DeBraS \cite{Municio:2016:DBS:2980137.2980143} avoids pro-actively cell overlapping and reduces internal interference by allowing nodes to share scheduling information. The cost for collision reduction and throughput improvement by DeBraS for dense networks is a higher energy consumption. 
	The algorithm proposed in \cite{Hwang:2017:DSA:3066575.3066845} allows every sensor node to compute its time-slot schedule in a distributed manner.
	A scheme called Reliable, Efficient, Fair and Interference-Aware Congestion Control (REFIACC) takes into account the heterogeneity in link interference and capacity when constructing the scheduling send policy in order to reach maximum fair throughput in wireless sensor networks \cite{Kafi20171}.
	The authors in \cite{7478619} proposed a "housekeeping" mechanism which detects scheduled collisions and reallocates each colliding cell to a different position in the schedule. A distributed cell-selection algorithm for reducing scheduling errors and collisions is proposed in \cite{Duy201780}. Scheduling Function Zero (SF0) adapts dynamically the number of reserved cells between neighboring nodes based on the application's bandwidth requirements and the network conditions \cite{dujovne-6tisch-6top-sf0-01}. SF0 uses Packet Delivery Rate (PDR) statistics to reallocate cells when the PDR of one or more cells is much lower than the average.
	Readers interested in an extended literature survey about scheduling algorithms in IEEE 802.15.4e are referred to \cite{DeGuglielmo20161}.
	
	The aforementioned scheduling algorithms are optimized for a specific performance metric such as energy consumption, reliability, latency, or throughput. In this paper, we propose a distributed link scheduling algorithm called \emph{Local Voting} (LV) that provides simultaneously low end-to-end latency and high reliability but on a significantly lower energy cost compared to existing algorithms in the literature. LV stems from the finding that the shortest delivery time is obtained when the load is equalized throughout the network \cite{7047923,8091101}.
	The performance of LV is studied through extensive simulation results in the 6TiSCH simulator \cite{simul}.
	
	The rest of the paper is organized as follows. In Section \ref{model}, the network model and problem are formulated. Section \ref{lv} presents Local Voting algorithm. Section \ref{perf} evaluates the performance of LV, and Section \ref{conc} concludes the paper.

	\section{Network Model and Problem Formulation}\label{model}
	Our model considers a 6TiSCH network which has built a tree routing topology by Routing Protocol for Low-Power and Lossy Networks (RPL) \cite{rfc6550}.
	The communication in the network can be modeled by a graph ${\cal G}=(V,E)$ where $V=\{n_i:  0 \leq i < N\}$ is the set of all nodes and $E$ is the set of edges that represent the communication symmetric links between the nodes.
	Data is gathered over a tree structure ${\cal G}_T=(V_T,E_T)$ rooted at the sink node $n_0$ where $n_0\in V_T, V_T\subseteq V$, and $E_T\subseteq E$. 
	Without loss of generality, we consider a single-sink model although the algorithm works in a model with multiple sinks. 
	We assume that all nodes are synchronized, and each node has a single half-duplex radio transceiver. 
	We propose a {\it{link scheduling}} algorithm where a link $(i, j)$ is a pairwise assignment of a directed communication between a pair of nodes $(n_i, n_j)$, where $i\neq j$, in a specific time slot within a given frame and a channel. 
	
\begin{figure}
		\centering
		\includegraphics[width=8.7cm,height=4.1cm]{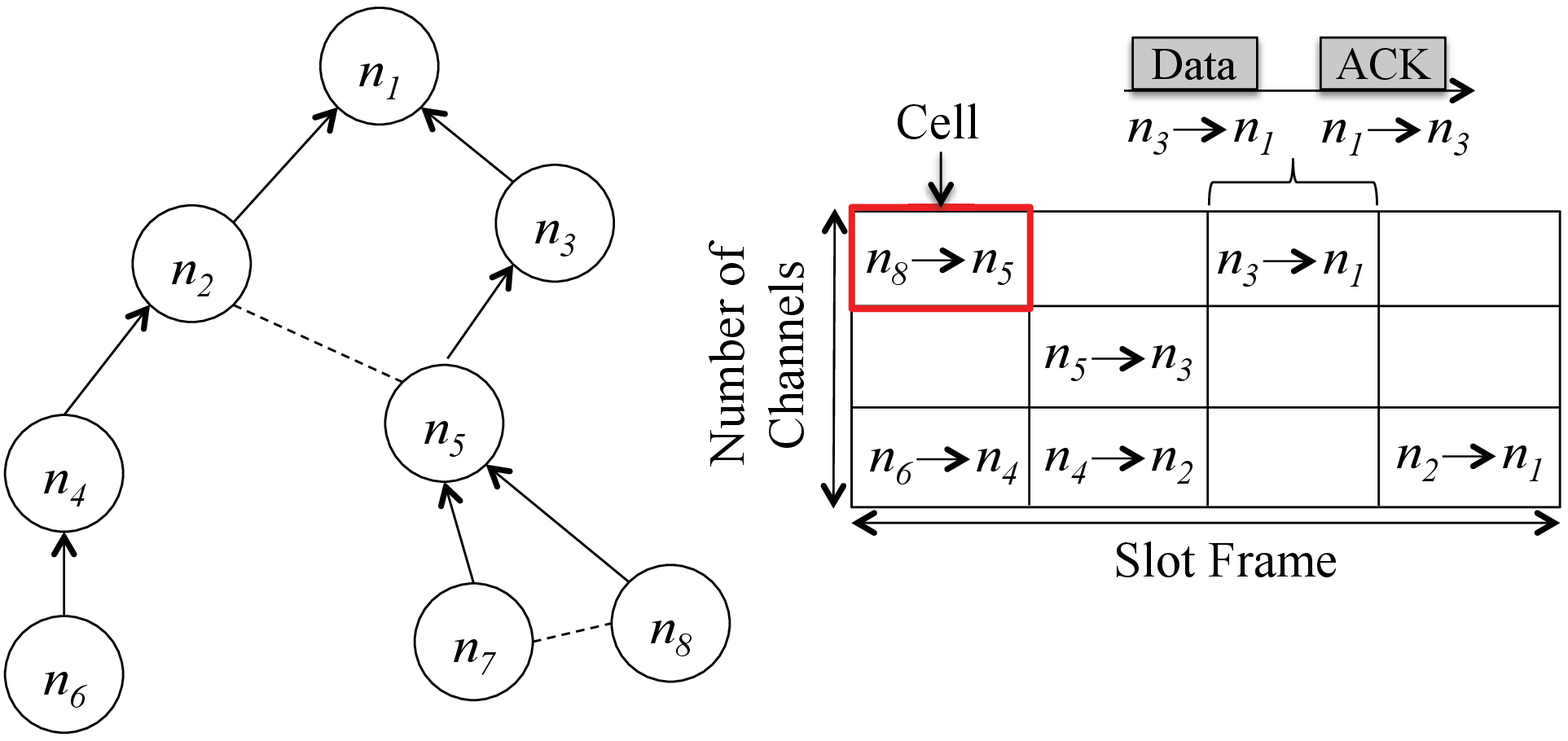}
		\caption{TSCH schedule for the presented topology where solid lines represent connection between nodes based on RPL and dashed lines represent possible communication between nodes.}
		\label{schedule}
	\end{figure}

	Time is divided into slot frames where each frame $f$ consists of equal number of $S$ time slots $f=\{0,\ldots, S-1\}$ with the same duration. A time slot $t$ is long enough for a MAC frame of maximum size to be sent from node $n_i$ to node $n_j$ and for node $n_j$ to reply. This is represented in Fig. \ref{schedule} for the node pair $(n_3, n_1)$. 
	The resource allocation in a TSCH network is controlled by a TSCH schedule that allocates cells for node communication. One example of a schedule with 4 time slots and 3 channels is given in Fig. \ref{schedule}. 
	A cell represents a unit of bandwidth that is allocated based on a decision by a centralized or a distributed scheduling algorithm. 
	Each cell is a pair of slot and channel offset coordinates assigned to a given link. The slot offset is equal to time slot $t$ while the channel offset \emph{chOf} is translated into a frequency using a function defined in the standard \cite{rfc7554}. The number of channel offsets is equal to the number of available frequencies $0\leq \emph{chOf} < M$.
	A TSCH schedule instructs node $n_i$ what to do in a specific time slot and frequency: transmit, receive, or sleep.  
	The cell assigned to link $(i, j)$ in slot offset $t$ and channel offset $\emph{chOf}$ is denoted by $c^{(t, \emph{chOf})}_{(i, j)}$ where 
	\begin{equation} 
	\label{cell}
	c^{(t, \emph{chOf})}_{(i, j)} = \left\{
	\begin{array}{rl}
	1, &n_i \mbox{ transmits and } n_j \mbox{ receives in } t \mbox{ and } \emph{chOf};\\
	0, &n_i \mbox{ and } n_j \mbox{ sleep in } t \mbox{ and } \emph{chOf}.
	\end{array} \right.
	\end{equation}
	
	Each scheduled cell is an opportunity for node $n_i$ to communicate with its one-hop neighbor $n_j$ where $n_j \in N_i^{(1)}$ and $N_i^{(1)}$ denotes the one-hop neighborhood of node $n_i$. We consider an interference model where two nodes are one-hop neighbors as long as their Packet Delivery Rate (PDR) is larger than 0. 
	
	The 6top sublayer qualifies each cell as either a hard or a soft cell. A soft cell can be read, added, deleted, or updated by the 6top sublayer, while a hard cell is read-only for the 6top sublayer. In the context of the proposed algorithm, all reallocated cells are soft cells. 
	
	The role of the scheduler is to ensure that there are enough resources to satisfy the needs of the applications (traffic load, end-to-end delay, reliability).
	The proposed scheduling algorithm must satisfy the following communication conditions:
	\begin{enumerate}
		\item Multi-point to point communication where data is generated only by source nodes $n_i$, where $n_i \in V_T$, and it is gathered at the sink node $n_0$.
		\item  The communication is half-duplex, thus, each node cannot transmit and receive simultaneously on the same channel. 
		\item Nodes $n_i$ and $n_j$ from the pair $(n_i, n_j)$ transmit and receive in the same cell, i.e. $(t, \emph{chOf})$, respectively.
		\item \textit{Collision-free} communication: A cell with coordinates $(t, \emph{chOf})$ is allocated to link $(i, j)$ such that exactly one of the neighbors, i.e. node $n_i$, of the receiving node $n_j$ should transmit in slot offset $t$ and channel offset $\emph{chOf}$, and the other neighbors $n_l$ of the receiving node $n_j$, where $n_l \in N_{j}^{(1)}$ and $n_l\neq n_i$, might receive in slot offset $t$ and channel offset $\emph{chOf}$.
	\end{enumerate}
	In general, to prevent collisions between pairs of links $(i, j)$ and $(l, k)$, the following \textit{collision-free} constraints are verified:
	
	\begin{equation} \label{collision-free1}
	c^{(t, \emph{chOf1})}_{(i, j)} c^{(t, \emph{chOf2})}_{(l, k)} = 0, \{i,j\}\cap\{k,l\}\neq \emptyset, n_k \in N_{i}^{(1)}, n_l \in N_j^{(1)}
	\end{equation}
	and
	\begin{equation} \label{collision-free2}
	c^{(t, \emph{chOf})}_{(i, j)} c^{(t, \emph{chOf})}_{(l, k)} = 0, n_k \in N_{i}^{(1)}, n_l \in N_j^{(1)}.
	\end{equation} 
	
	Eq.~(\ref{collision-free1}) indicates that the communication is half-duplex, which is also known as a \emph{primary conflict constraint}. 
	Namely, a node cannot transmit and/or receive two packets at the same time slot $t$, even not on different channels \emph{chOf1} and \emph{chOf2}. 
	
	Eq.~(\ref{collision-free2}) indicates the interference constraint, also known as a \emph{secondary conflict constraint}. It stems from the fact that a receiver cannot decode 
	an incoming packet in a channel \emph{chOf}, if another node in its neighborhood is also transmitting at the same channel \emph{chOf} at the same time slot $t$. Hence, a node is not allowed to receive more than one transmission simultaneously.

	
	\section{Local Voting Cell Allocation}\label{lv}
	
	Each source node $n_i$, where $n_i \in V_T$ and $n_i\neq n_0$, has a queue with packets to be transmitted to a specific one-hop neighbor. The internal scheduling on the queue is first-come-first-serve. A cell is allocated to link $(i, j)$ so that node $n_i$ transmits a packet to $n_j$ as it presented in Eq. (\ref{cell}).  
	
	The state of each pair of nodes $(n_i, n_j)$, where $n_j \in N_i^{(1)}$, at the beginning of frame $f+1$ is described by two characteristics:
	\begin{itemize}
		\item $q_{(i, j)}^{f+1}$ is the number of packets (queue length) that node $n_i$ has to transmit to node $n_j$ at slot frame $f+1$;
		\item $p_{(i, j)}^{f}$ is the number of cells allocated to link $(i, j)$ at the previous slot frame $f$, i.e. $p_{(i, j)}^{f} = \sum\limits_{t=0}^{S-1}  c_{(i, j)}^{(t, \emph{chOf})}$.
	\end{itemize}
	There is no sum over the channels in the equation for calculating $p_{(i, j)}^{f}$ due to the fact that each node has a single transceiver.
	
	The dynamics of each link $(i, j)$ are calculated as:
	\begin{eqnarray}
	\label{dynamics}
	\begin{aligned}
	q_{(i, j)}^{f+1} &= \max\{0, q_{(i, j)}^{f} - p_{(i, j)}^{f+1} \} + z_{(i, j)}^{f},  \\
	p_{(i, j)}^{f+1} &= p_{(i, j)}^{f} + u_{(i, j)}^{f+1},
	\end{aligned}
	\end{eqnarray}
	where 
	\begin{itemize}
		\item $z_{(i, j)}^{f}$ is the number of new packets received from upper layers or from neighboring nodes of node $n_i$ with a next-hop destination equal to node $n_j$ at frame $f$;
		\item $u_{(i, j)}^{f+1}$ is the number of cells that are added or released 
		to link $(i, j)$ at frame $f+1$ due to LV. 
	\end{itemize}
	Note that number of cells for $p_{(i, j)}^{f}$ and $u_{(i, j)}^{f+1}$ is also equal to the number of time slots needed, since all transmissions from the same source are in primary conflict.
	
	In the following part we explain LV and the way how $u_{(i, j)}^{f+1}$ is calculated. LV triggers the 6top sublayer to add and release cells to link $(i, j)$ at frame $f+1$ for $u_{(i, j)}^{f+1}>0$ and $u_{(i, j)}^{f+1}<0$, respectively.
	
	The objective of the proposed LV algorithm is to schedule link transmissions in such a way that the minimum maximal (min-max) link delay is achieved. The algorithm stems from the finding that the shortest delivery time is obtained when the load is equalized throughout the network. We refer to Lemma 1 and Corollary 1 of \cite{7047923} for showing that the minimum expected nodal delay is achieved when the load in the network is equalized on nodes. Later, it was proved in \cite{8091101} that the network system converges asymptotically towards optimal node scheduling. 
	Note that references \cite{7047923,8116537,8091101} consider a node scheduling problem, while in this paper we consider a link scheduling problem with multiple channels. Proving the optimality of LV as a link scheduling algorithm is a future work. 
	
	\begin{table*}[t]
		\centering
		\caption{An example of the evolution of Local Voting algorithm applied for the network in Fig. \ref{schedule}}\label{lvSchedule}
		\resizebox{\linewidth}{!}{
			\begin{tabular}{ |c|c|c|c|c|c|c|c|c|c|c|c|c|c|c|c|c|c|c|c|c|c|c|c|c|c|c|c|c|c|c|c| } 
				\hline
				$(i,j)$ & \multicolumn{4}{|c|}{(6, 4)} 	& \multicolumn{4}{|c|}{(3, 1)} & \multicolumn{4}{|c|}{(2, 1)} & \multicolumn{4}{|c|}{(7, 5)}  & \multicolumn{4}{|c|}{(4, 2)} & \multicolumn{4}{|c|}{(8, 5)} & \multicolumn{4}{|c|}{(5, 3)} \\
				\hline
				$f$ & $p$ & $q$ & $x$ & $u$ & $p$ & $q$ & $x$ & $u$ & $p$ & $q$ & $x$ & $u$ & $p$ & $q$ & $x$ & $u$ & $p$ & $q$ & $x$ & $u$ & $p$ & $q$ & $x$ & $u$ & $p$ & $q$ & $x$ & $u$\\ 
				\hline
				0 & 0 & 10 & NA & 3 & 0 & 20 & NA & 7 & 0 & 5 & NA & 1 & 0 & 25 & NA & 7 & 0 & 45 & NA & 11 & 0 & 7 & NA & 2 & \textbf{\textcolor{red}{0}} & \textbf{\textcolor{red}{14}} & \textbf{\textcolor{red}{NA}} & \textbf{\textcolor{red}{3}}\\ 
				\hline
				1  & 3 & 7 & 3 & -1 & 7 & 16 & 3 & -3 & 1 & 15 & 16 & 2 & 7 & 18 & 3 & -2 & 11 & 37 & 4 & -2 & 2 & 5 & 3 & 0 & \textbf{\textcolor{red}{3}} & \textbf{\textcolor{red}{20}} & \textbf{\textcolor{red}{7}} & \textbf{\textcolor{red}{2}}\\ 
				\hline
				2 & 2 & 5 & 3 & 0 & 4 & 17 & 5 & 0 & 3 & 21 & 8 & 1 & 5 & 13 & 3 & -1 & 9 & 30 & 4 & -2 & 2 & 3 & 2 & -1 & 5 & 22 & 5 & 0\\ 
				\hline
				3  & 2 & 3 & 2 & -1 & 4 & 18 & 5 & 0 & 4 & 24 & 7 & 1 & 4 & 9 & 3 & -1 & 7 & 25 & 4 & 0 & 1 & 2 & 3 & 0 & 5 & 22 & 5 & 1\\ 
				\hline
				4 & 1 & 2 & 3 & 0 & 4 & 20 & 6 & 0 & 5 & 26 & 6 & 1 & 3 & 6 & 3 & -1 & 7 & 19 & 3 & -1 & 1 & 1 & 2 & -1 & 6 & 20 & 4 & 0\\ 
				\hline
				5  & 1 & 1 & 2 & 0 & 4 & 22 & 6 & 1 & 6 & 26 & 5 & 0 & 2 & 4 & 3 & 0 & 6 & 14 & 3 & -1 & 0 & 1 & NA & 0 & 6 & 16 & 3 & -1\\ 
				\hline
				6 & 1 & 0 & 1 & -1 & 5 & 22 & 5 & 0 & 6 & 25 & 5 & 1 & 2 & 2 & 2 & -1 & 5 & 10 & 3 & -1 & 0 & 1 & NA & 1 & 5 & 13 & 3 & 0\\ 
				\hline
				7  & 0 & 0 & NA & 0 & 5 & 22 & 5 & 1 & 7 & 22 & 4 & 0 & 1 & 1 & 2 & 0 & 4 & 6 & 2 & -1 & 1 & 0 & 1 & -1 & 5 & 10 & 3 & -1\\ 
				\hline
				8  & 0 & 0 & NA & 0 & 6 & 20 & 4 & 1 & 7 & 18 & 3 & 0 & 1 & 0 & 1 & -1 & 3 & 3 & 2 & -1 & 0 & 0 & NA & 0 & 4 & 7 & 2 & 0\\ 
				\hline
				9 & 0 & 0 & NA & 0 & 6 & 18 & 4 & 2 & 7 & 13 & 2 & -1 & 0 & 0 & NA & 0 & 2 & 1 & 1 & -1 & 0 & 0 & NA & 0 & 4 & 3 & 1 & -2\\ 
				\hline
				10  & 0 & 0 & NA & 0 & 7 & 13 & 2 & 2 & 6 & 8 & 2 & 0 & 0 & 0 & NA & 0 & 1 & 0 & 1 & -1 & 0 & 0 & NA & 0 & 2 & 1 & 1 & -1\\ 
				\hline
				11 & 0 & 0 & NA & 0 & 8 & 6 & 1 & 3 & 6 & 2 & 1 & -2 & 0 & 0 & NA & 0 & 0 & 0 & NA & 0 & 0 & 0 & NA & 0 & 1 & 0 & 1 & -1\\ 
				\hline
				12  & 0 & 0 & NA & NA & 11 & 0 & 1 & NA & 4 & 0 & 1 & NA & 0 & 0 & NA & NA & 0 & 0 & NA & NA & 0 & 0 & NA & NA & 0 & 0 & NA & NA\\ 
				\hline
			\end{tabular}
		} 
	\end{table*}
	
	The load of link $(i, j)$ at frame $f$ is defined as the ratio of the queue length $q_{(i, j)}^{f}$ over the number of allocated cells $p_{(i, j)}^{f}$ as follows:
	\begin{equation} \label{load}
	x_{(i, j)}^{f} = \left\{
	\begin{array}{rl}
	\left[ \cfrac{q_{(i, j)}^{f}}{p_{(i, j)}^{f}} +0.5 \right], &\mbox{ if } q_{(i, j)}^{f} > 0, \\
	0, &\mbox{ if } q_{(i, j)}^{f}= 0,
	\end{array} \right. 
	\end{equation}
	where
	$\left[ \cdot \right]$ is the round function
	(rounds a real number to the nearest integer).
	
	In order to semi-equalize or balance the load in the network, neighboring links can exchange cells as long as Eq. (\ref{collision-free1}) and Eq. (\ref{collision-free2}) are satisfied. The set $N_{i,j}$ contains all links that could potentially interfere with link $(i,j)$. This means that 
	$$
	(l,k) \in N_{i,j} \mbox{ iff } n_k \in N^{(1)}_i \lor n_l \in N^{(1)}_j.
	$$
	The value of $u_{(i, j)}^{f+1}$ is calculated as:
	\begin{equation} \label{u}
	{u}_{(i, j)}^{f+1} = \left[ \cfrac{q_{(i, j)}^{f+1} \times S}{q_{(i, j)}^{f+1}+\sum_{(l,k) \in N_{i,j}}{w_{(i,j,l,k)} \times q_{(l, k)}^{f+1}}} \right] - p_{(i, j)}^{f},
	\end{equation}
	where 
	\begin{equation}\label{w}
	w_{(i,j,l,k)} = \left\{
	\begin{array}{rl}
	1, &\mbox{ if } \{i,j\}\cap\{k,l\}\neq \emptyset, \\
	\sfrac{1}{M}, &\mbox{ othewise. }
	\end{array} \right. 
	\end{equation}
	
	Eq. (\ref{u}) and Eq. (\ref{w}) are explained as follows: the value in the round function in Eq. (\ref{u}) is 
	the number of cells allocated to link $(i,j)$ at frame $f+1$. As we can see from the term $q_{(i, j)}^{f+1}$, the number of allocated cells is proportional to the queue length within the neighborhood of link $(i,j)$, so it leads to
	semi-equal load between the neighboring links. 
	Also, we scale to the total number of time slots that are needed to transmit all 
	queued packets in the neighborhood of 
	link $(i,j)$, so that the total number of time slots in the neighborhood is 
	equal to
	the number of time slots in the frame. 
	The weight $w_{(i,j,l,k)}$ is used to capture the difference between a primary and a secondary
	conflict. In the first case, since all channels are unavailable to the link, the 
	value is one, but in the second case, since only one of the available channels
	is blocked, the value is $1/M$.
	\begin{example}
		To illustrate the proposed LV algorithm, consider the network given in Fig. \ref{schedule}. Assume that the total number of cells in the schedule is 75 where the number of channels is $M=5$ and the number of time slots per slot frame is $S=15$. 
		Assume that the initial queue lengths are: $q_{(6,4)}^0 = 10, \; q_{(3,1)}^0 = 20, \; q_{(2,1)}^0 = 5, \; q_{(7,5)}^0 = 25, \; q_{(4,2)}^0 = 45, \; q_{(8,5)}^0 = 7, \;$and $q_{(5,3)}^0 = 14$. These values correspond to the values of $q$ for $f=0$ for each of the links given in Table \ref{lvSchedule}.
		
		We next show how the values presented in red color for the link $(5, 3)$ in Table \ref{lvSchedule} are calculated for the first two frames, i.e. $f=0$ and $1$.
		The queue length $q$ for $f=0$ is equal to $14$. In the beginning, no slots are allocated to link $(5,3)$. Hence, $p=0$ and we do not calculate the load $x$ since it cannot be defined for $0$ slot allocation. The $u$ value is calculated with Eq. (\ref{u}). The link $(5,3)$ is in a primary conflict with the links $(7,5), (8,5),$ and $(3,1),$ and the value of $w_{(i,j,l,k)}$ for these links is 1. On the other hand, the link $(5,3)$ is in a secondary conflict with the link $(4,2)$ and the value of $w_{(5,3,4,2)}$ is 1/5. It follows that 
		\begin{eqnarray}
		\nonumber
		u_{(5,3)}^0 = & \left[ \cfrac{14 \times 15}{14 + 1 \times (25+7+20) + \sfrac{1}{5}\times 45} \right] - 0 & = 3.\\	
		\nonumber	
		\end{eqnarray}
		This means that LV triggers the 6top sublayer to allocate 3 cells to the link $(5,3)$. Following Eq. (\ref{dynamics}), the number of allocated cells $p$ and the queue length $q$ for $f=1$ become 3 and 20, respectively. Although 3 packets have been sent, still new packets have been received from the links $(7,5)$ and $(8,5)$. Therefore, the queue length for $f=1$ becomes $q=\max\{0,14-3\}+7+2=20$. The load $x$ is calculated as
		$$x_{(5,3)}^1 = \left[ \sfrac{20}{3}+0.5 \right] = 7.$$ 
		The value of $u$ is calculated in a similar way as it was presented for $f=0$ and so forth.
		As we can see from Table \ref{lvSchedule}, the load is equalized for all links in the 12-th frame.
	\end{example}
	
	\subsection{Local Voting Algorithm}
	Alg. 1 presents Local Voting.
	All links (edges) are examined sequentially at the beginning of each frame. The source node requests for cells, not the receiver. Since we consider a link scheduling scenario, the destination of each transmission is known during the scheduling phase. 
	Every link in the network that has a positive queue length calculates a value $u^{f+1}$ (given in Eq. (\ref{u})). 
	If node $n_i$ has packets to send to node $n_j$, the value of ${u}_{(i, j)}^{f+1}$ determines how many cells the link $(i,j)$ should ideally gain or release at slot frame $f+1$.
	If $u_{(i,j)}^{f+1}$ is a positive value, then LV asks from the 6top sublayer to add cells to link $(i,j)$. Otherwise, if $u_{(i,j)}^{f+1}$ is a negative value, then LV requests from the 6top sublayer to release $u_{(i,j)}^{f+1}$ cells that have been allocated to $(i,j)$.
	The cell reallocation should not cause collisions with respect to Eq. (\ref{collision-free1}) and Eq. (\ref{collision-free2}). The collision-free constraint is implemented in 6top sublayer which is responsible for eventually reaching collision-free communication.
	On the other hand, if node $n_i$ does not have packets to send to destination $n_j$ and cells have been already allocated to link $(i,j)$ in the previous frame, then all allocated cells $p_{(i,j)}^{f}$ are released. In general, cells are removed from links with a lower load and are offered to links with a higher load.
	
	\begin{algorithm} \label{alg2}
		\floatname{algorithm}{Algorithm}
		\begin{algorithmic}
			\For {$(i, j) \in E$}   \Comment{Check for all outgoing links $(i,j)$ that originate at node $n_i$}
			\State{$qsum_{(i,j)}^{f+1} = q_{(i, j)}^{f+1}+\sum_{(l,k) \in N_{i,j}}{w_{(i,j,l,k)} \times q_{(l, k)}^{f+1}}$}
			\If {$qsum_{(i,j)}^{f+1} \neq 0$} \Comment {Are there packets in the neighborhood of link $(i,j)$ to be sent?}
			\State Calculate $u_{(i,j)}^{f+1} = \left[ \frac{q_{(i, j)}^{f+1} \times S}{qsum_{(i,j)}^{f+1}} \right] - p_{(i, j)}^{f} $ 
			\If {$u_{(i,j)}^{f+1} > 0$} \Comment{The link requests cells}
			\State{Request from 6top to add $u_{(i,j)}^{f+1}$ cells to link~$(i,j)$}
			\ElsIf { $u_{(i,j)}^{f+1} < 0$} \Comment{The link releases cells}
			\State{Request from 6top to delete $u_{(i,j)}^{f+1}$ cells from link~$(i,j)$}
			\EndIf
			\ElsIf { $p_{(i,j)}^{f} \neq 0$} \Comment{Are there cells allocated to a link with an empty queue?}
			\State Request from 6top to delete $p_{(i,j)}^{f}$ cells from ~ link~$(i,j)$ \Comment{Release the allocated cells}	
			\EndIf
			\EndFor
		\end{algorithmic}
		\caption{Local Voting}
	\end{algorithm}
	To summarize, LV requests from the 6top sublayer to add cells to link $(i,j)$ at slot frame $f+1$ when:
	\begin{itemize}
		\item Node $n_i$ has packets to send to node $n_j$ and the value of $u_{(i,j)}^{f+1}$ for link $(i,j)$ is positive which means that the link $(i,j)$ has a higher load than its neighbors.
	\end{itemize}
	LV requests from the 6top sublayer to release cells from link $(i,j)$ at slot frame $f+1$ when:	
	\begin{itemize}
		\item Node $n_i$ has packets to send to node $n_j$ and the value of $u_{(i,j)}^{f+1}$ is negative which means that the link $(i,j)$ has a lower load than its neighbors; or
		\item Node $n_i$ does not have packets to send to node $n_j$ and cells have been already allocated to link $(i,j)$.
	\end{itemize}

	\section{Performance Evaluation}\label{perf}

\begin{figure*}
	\centering
	\subfigure[Queue length=100, parent size=1]
	{\includegraphics[width=0.3293\textwidth]{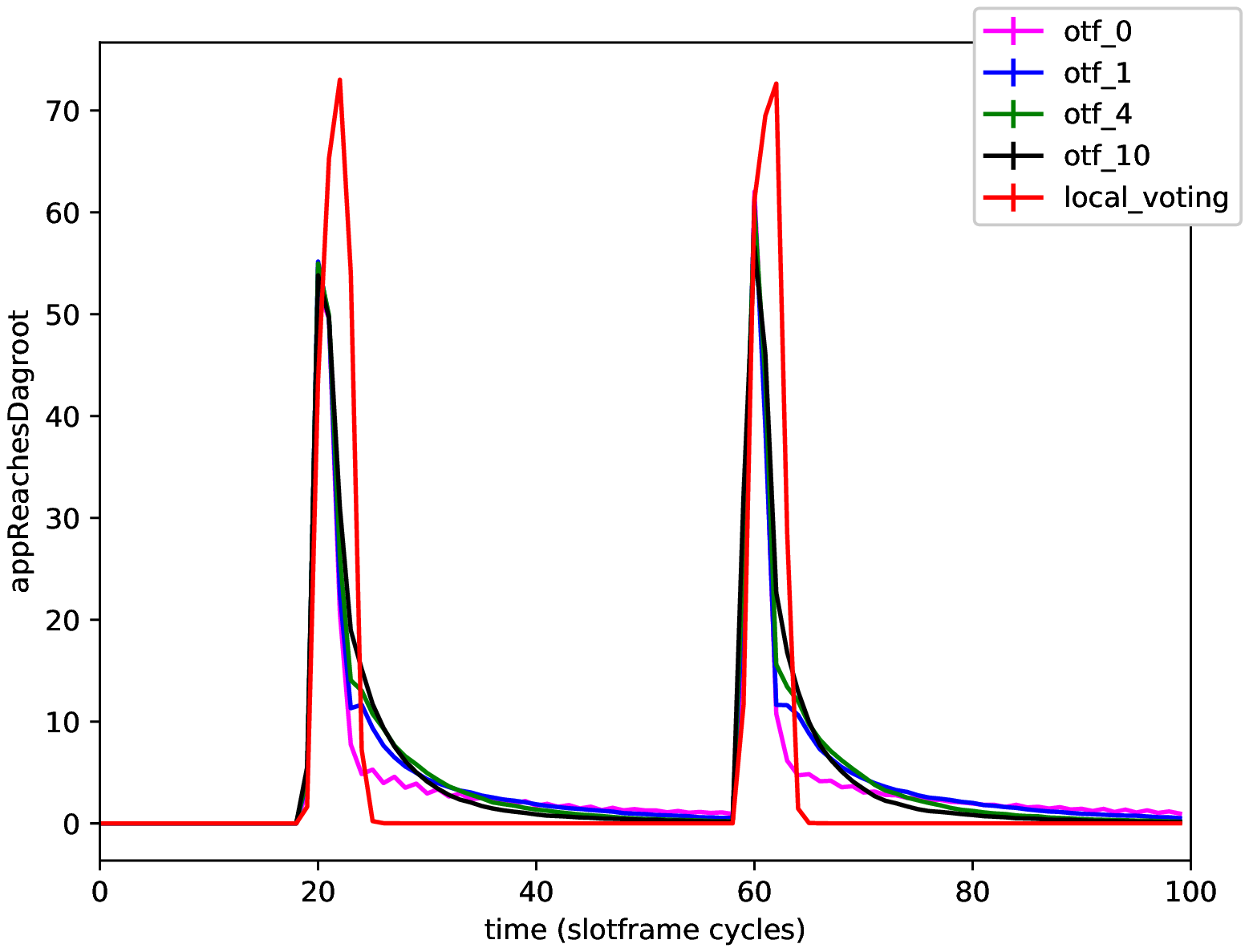}} 
	\hfill
	\subfigure[Queue length=100, parent size=2]
	{\includegraphics[width=0.3293\textwidth]{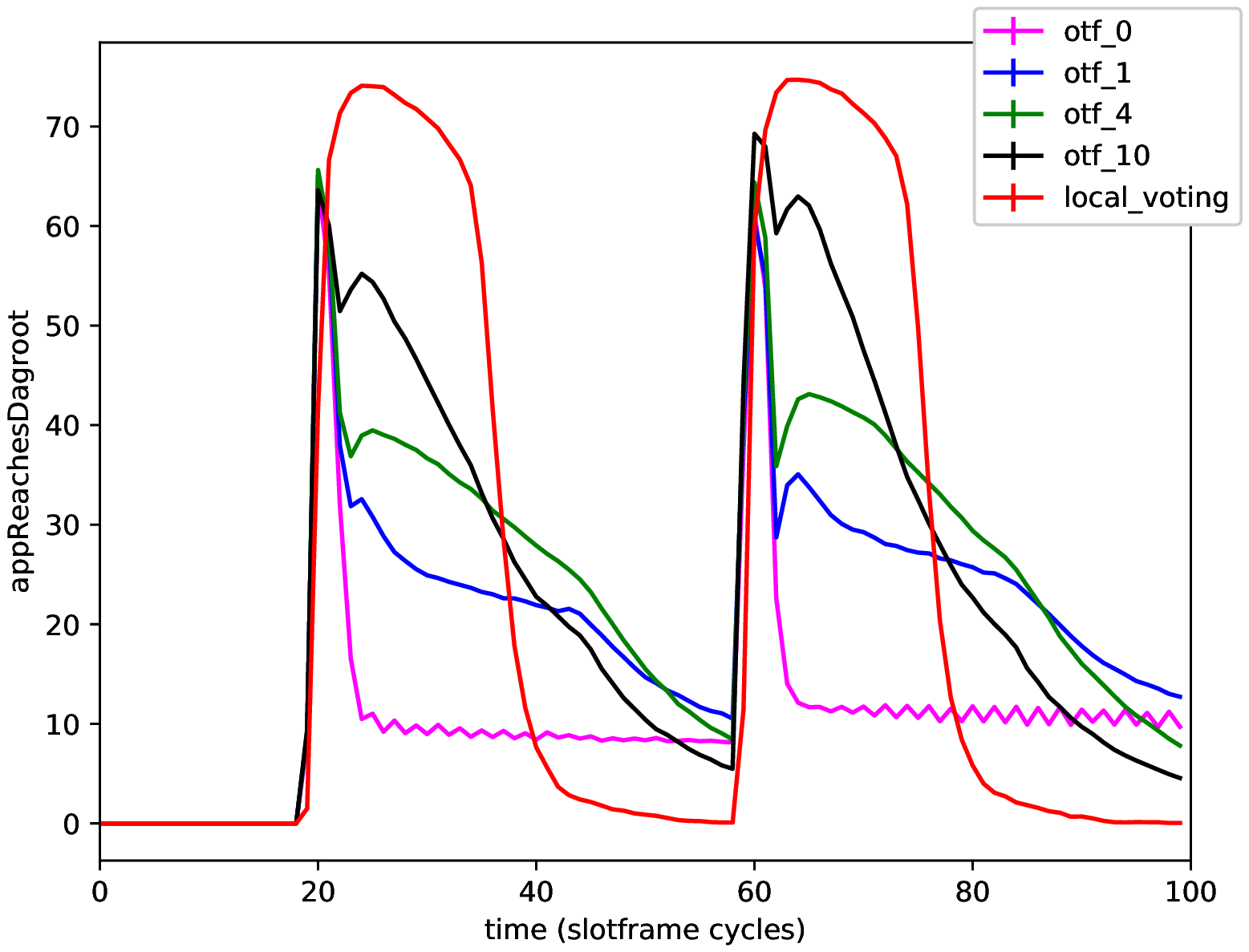}} 
	\hfill
	\subfigure[Queue length=100, parent size=3]
	{\includegraphics[width=0.3293\textwidth]{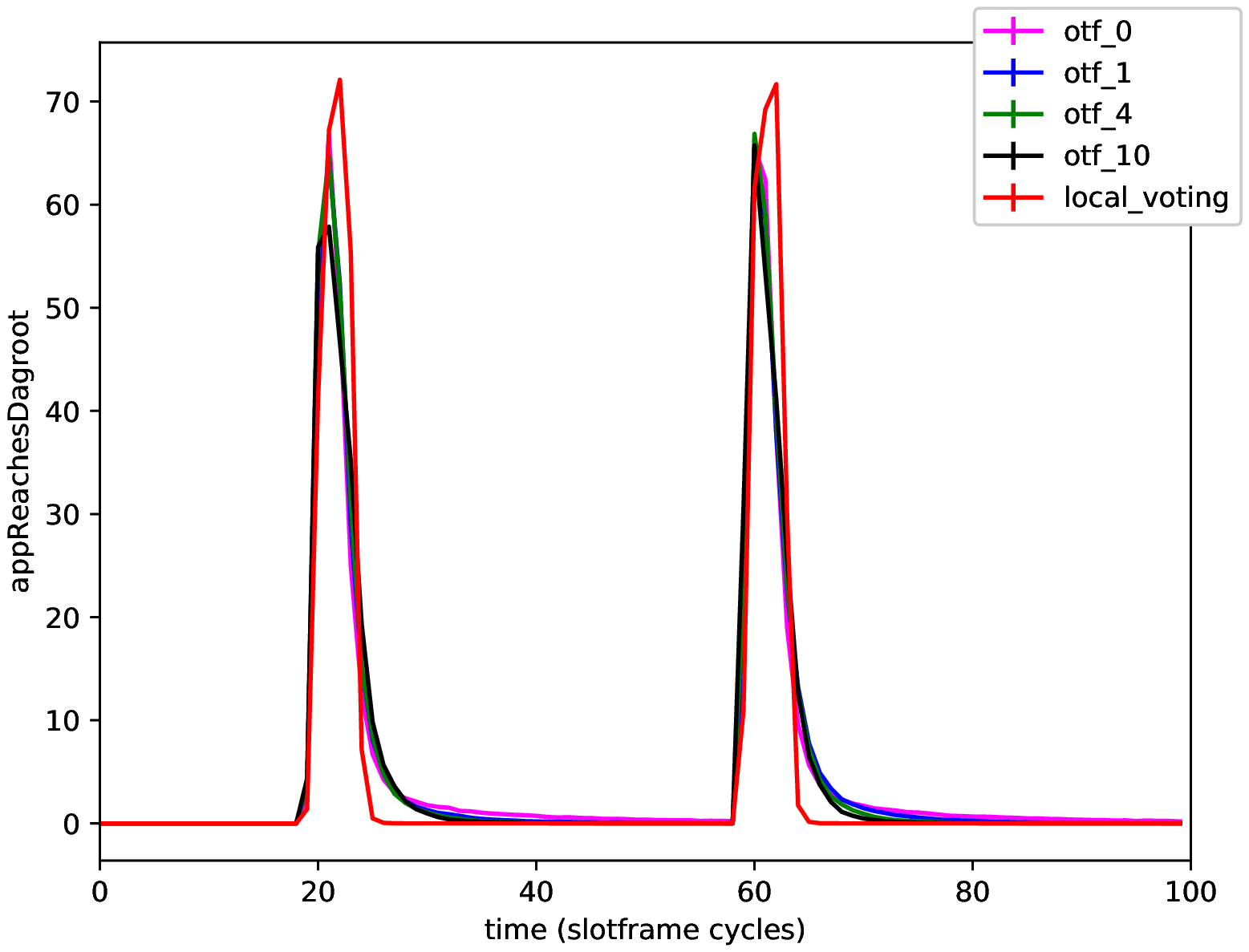}} 
	\caption{Number of packets that reach the root over time for queue length of 100 when each node generates (a) 5 packets per burst for 1 parent, (b) 25 packets per burst for 2 parents, and (c) 5 packets per burst for 3 parents.}
	\label{reliability}
\end{figure*}
\shorten{
	\begin{figure*}
		\centering
		\subfigure[Queue length=100, parent size=1]
		{\includegraphics[width=0.3293\textwidth]{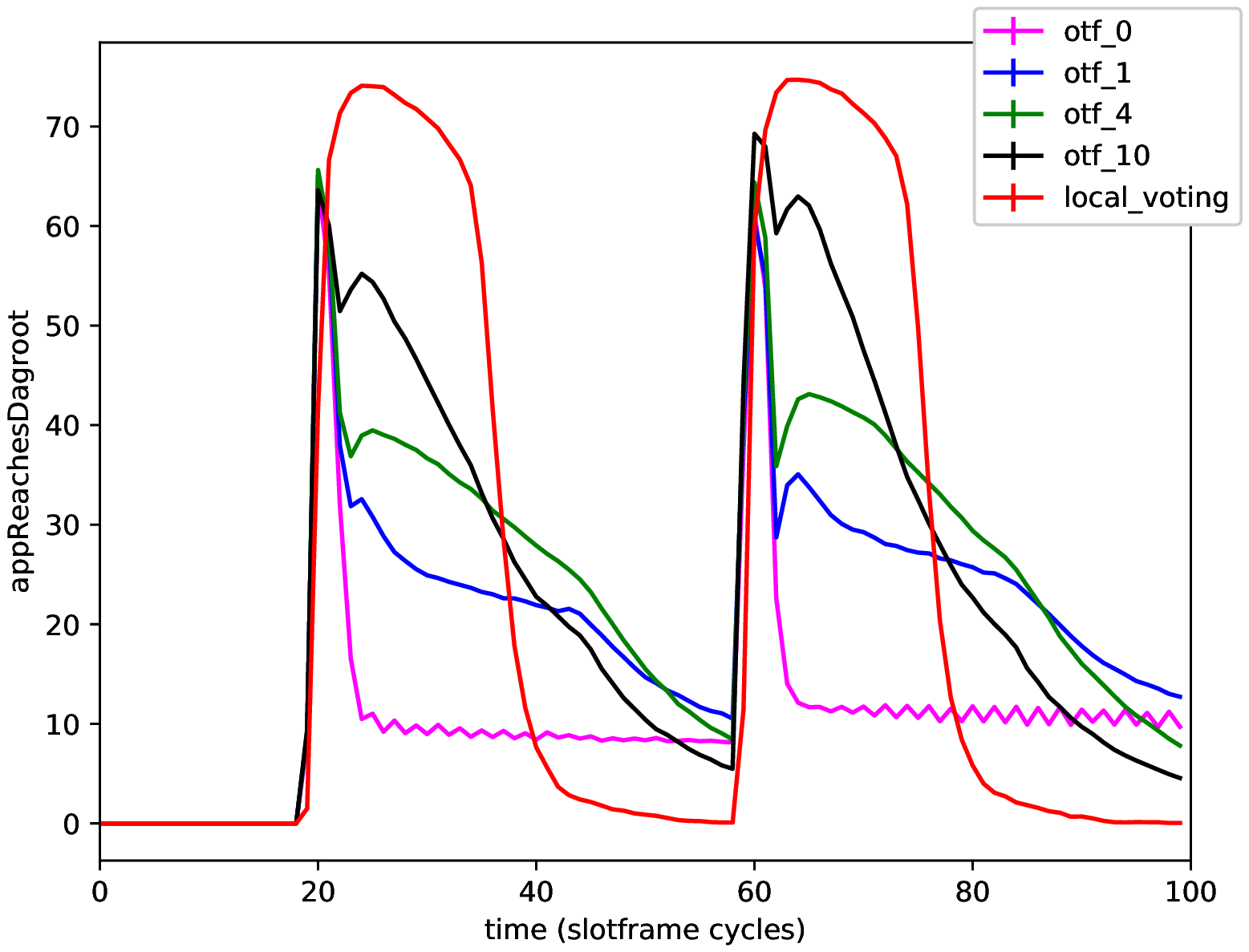}} 
		\hfill
		\subfigure[Queue length=100, parent size=2]
		{\includegraphics[width=0.3293\textwidth]{appReachesDagroot_vs_time_buf_100_par_2_pkt_25}} 
		\hfill
		\subfigure[Queue length=100, parent size=3]
		{\includegraphics[width=0.3293\textwidth]{appReachesDagroot_vs_time_buf_100_par_3_pkt_25}} 
		\caption{Number of packets that reach the root over time for queue length of 100 when each node generates 25 packets per burst for (a) 1 parent, (b) 2 parents, and (c) 3 parents.}
		\label{reliability25}
	\end{figure*}
}

\begin{figure*}[thpb]
	\centering
	\subfigure[Time for last packet to reach the root]
	{\includegraphics[width=0.3293\textwidth]{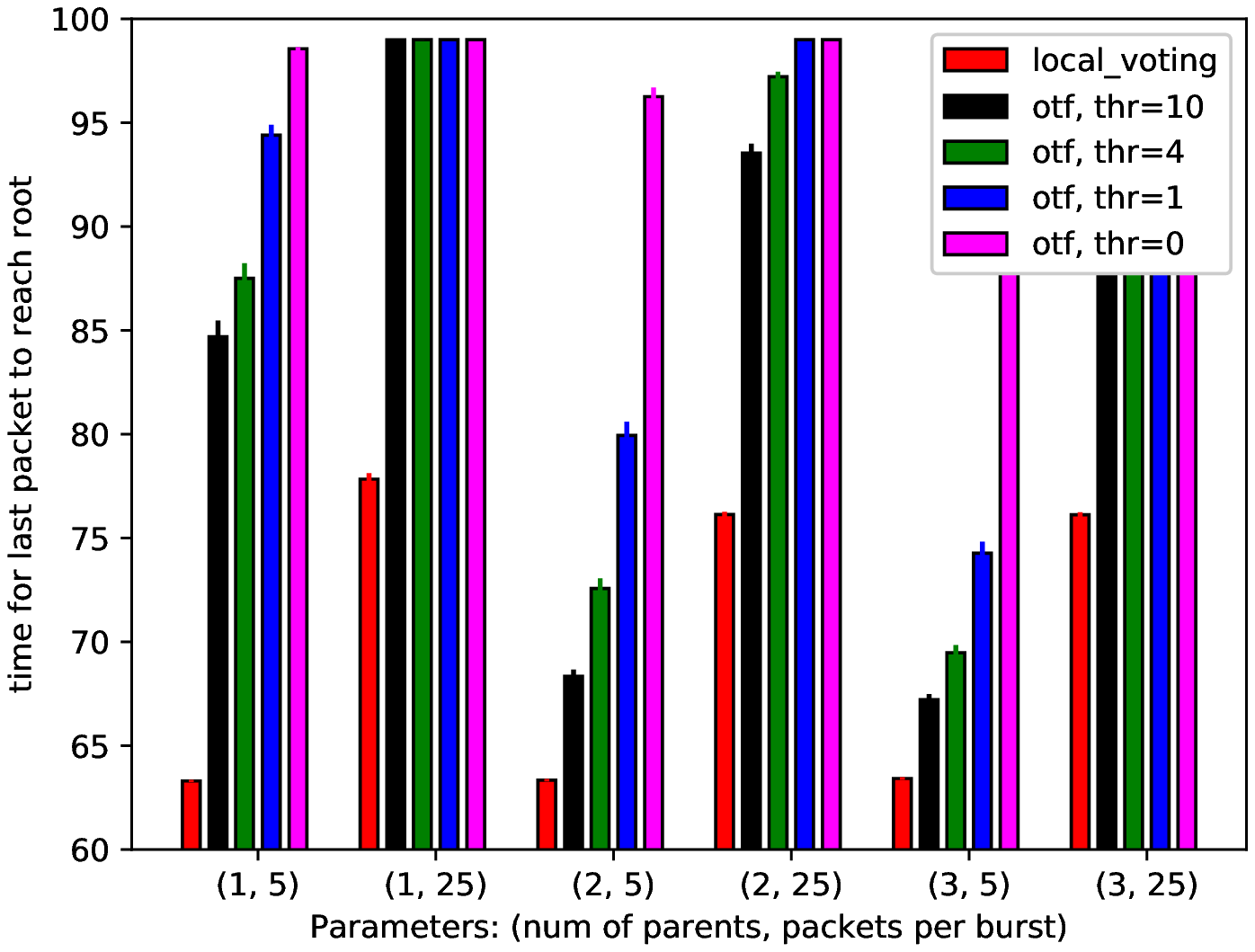}}
	\hfill
	\subfigure[Max end-to-end latency]
	{\includegraphics[width=0.3293\textwidth]{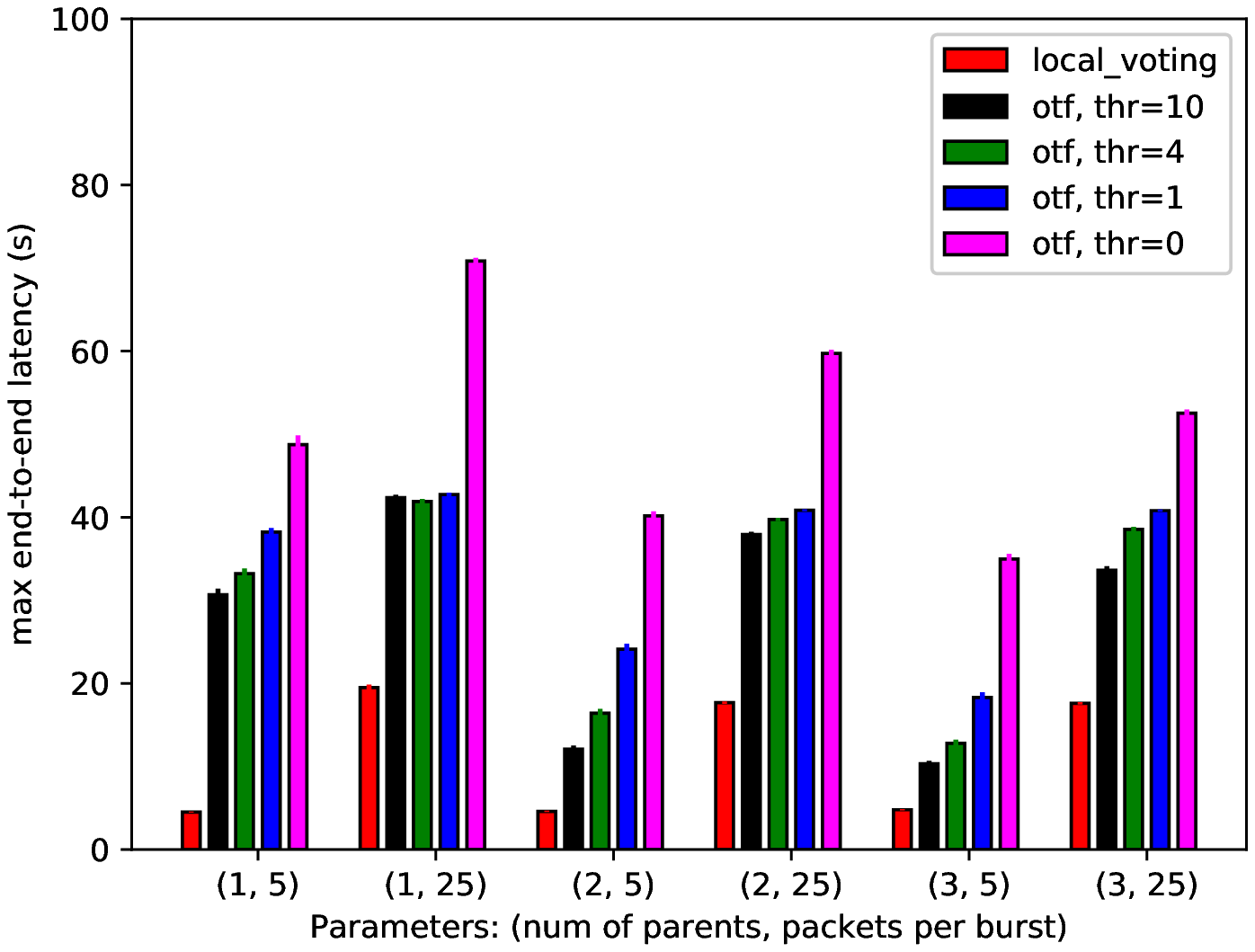}} 
	\hfill
	\subfigure[Charge Consumed]
	{\includegraphics[width=0.3293\textwidth]{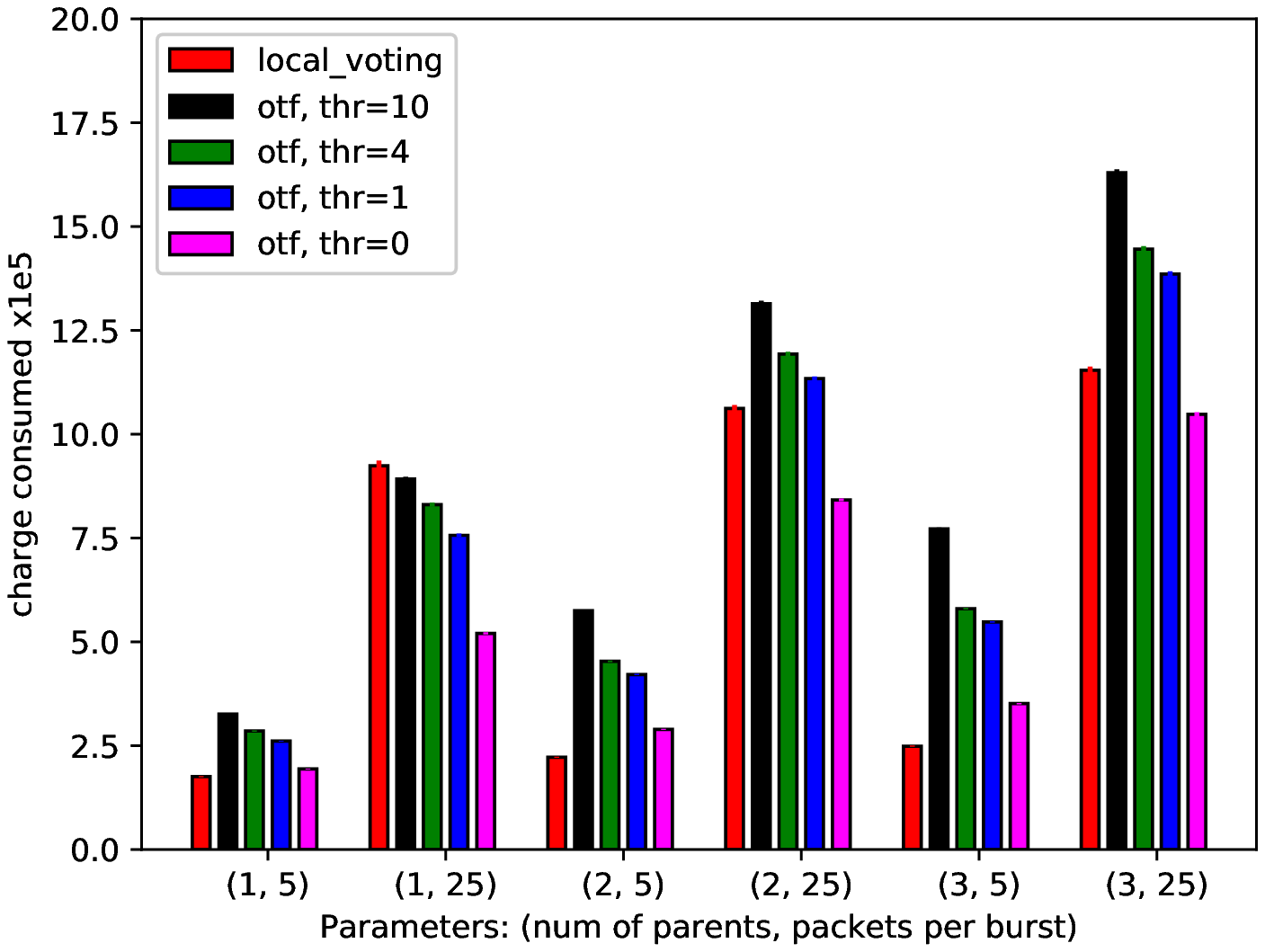}} 
	\caption{Aggregated results as a function of the number of RPL parents and
		packets per burst when the queue length is set to 100 packets and the confidence interval is 95\%.}
	\label{timeLast}
\end{figure*}


\shorten{
	\begin{figure*}
		\centering
		\subfigure[Queue length=100, parent size=1]
		{\includegraphics[width=0.3293\textwidth]{numRxCells_vs_time_buf_100_par_1_pkt_5}} 
		\hfill
		\subfigure[Queue length=100, parent size=2]
		{\includegraphics[width=0.3293\textwidth]{numRxCells_vs_time_buf_100_par_2_pkt_5}} 
		\hfill
		\subfigure[Queue length=100, parent size=3]
		{\includegraphics[width=0.3293\textwidth]{numRxCells_vs_time_buf_100_par_3_pkt_5}} 
		\caption{Number of allocated cells over time for queue length of 100 packets when each node generates 5 packets per burst for (a) 1 parent, (b) 2 parents, and (c) 3 parents.}
		\label{numRx}
	\end{figure*} 
}
\begin{figure*}
	\centering
	\subfigure[Queue length=100, parent size=1]
	{\includegraphics[width=0.3293\textwidth]{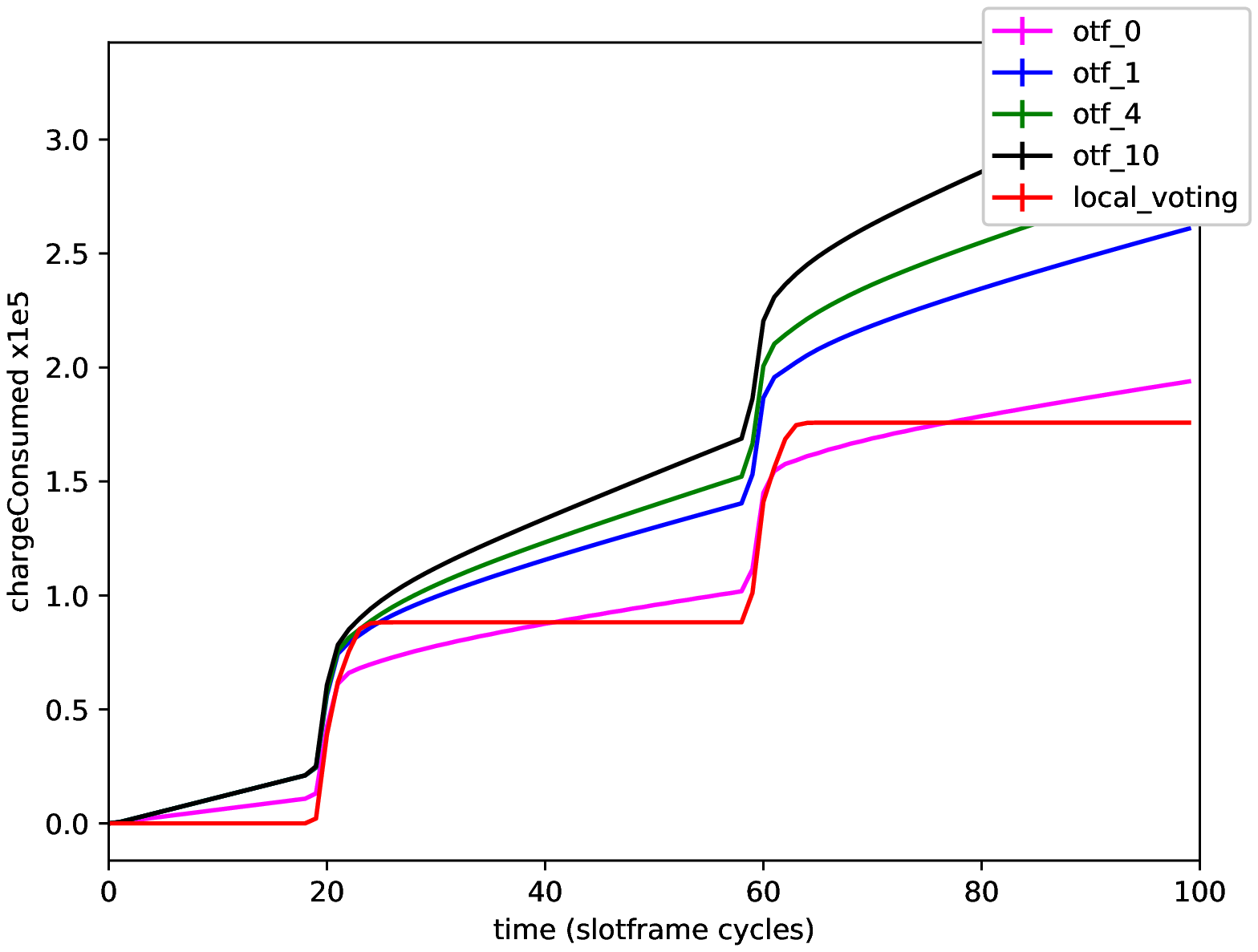}} 
	\hfill
	\subfigure[Queue length=100, parent size=2]
	{\includegraphics[width=0.3293\textwidth]{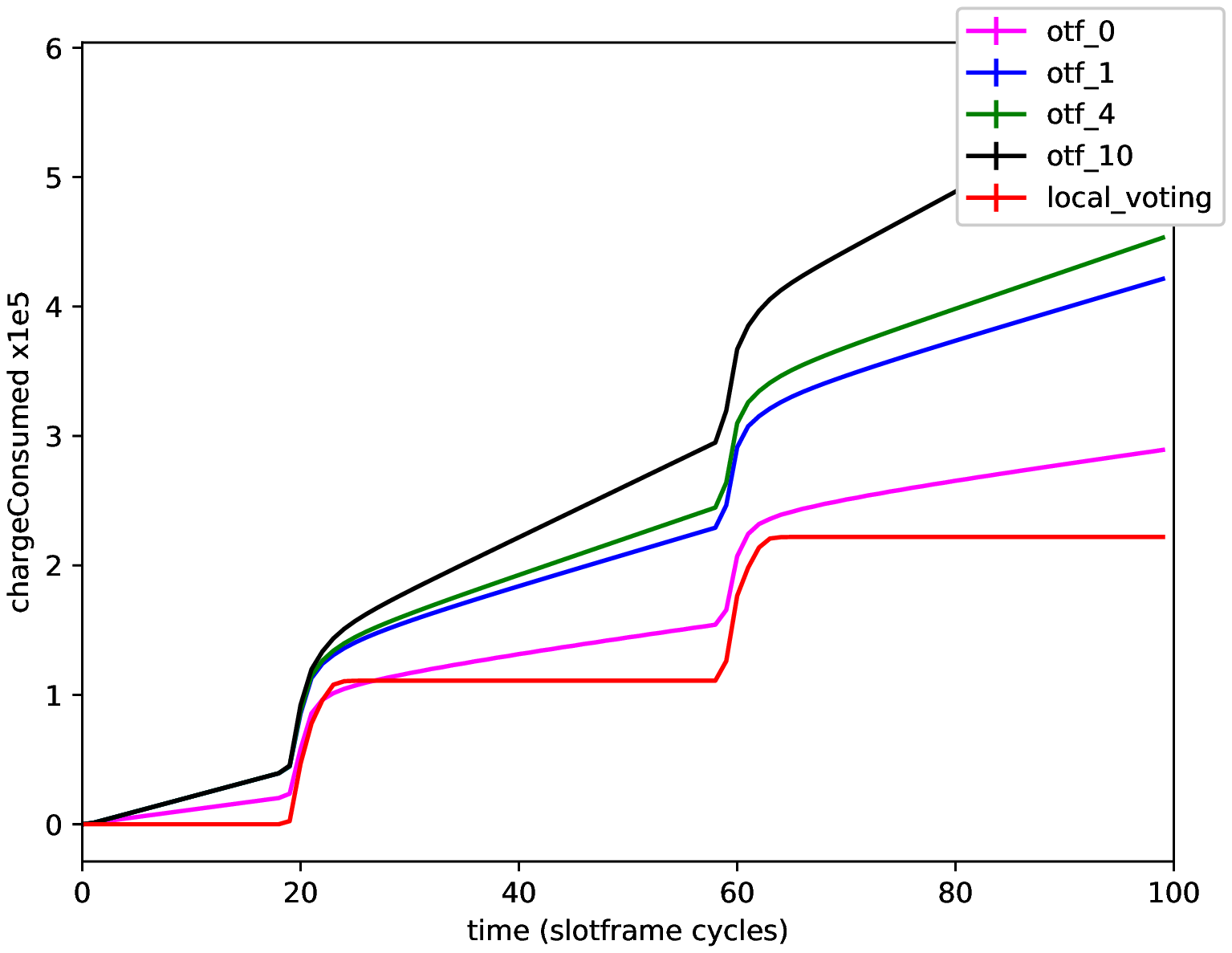}} 
	\hfill
	\subfigure[Queue length=100, parent size=3]
	{\includegraphics[width=0.3293\textwidth]{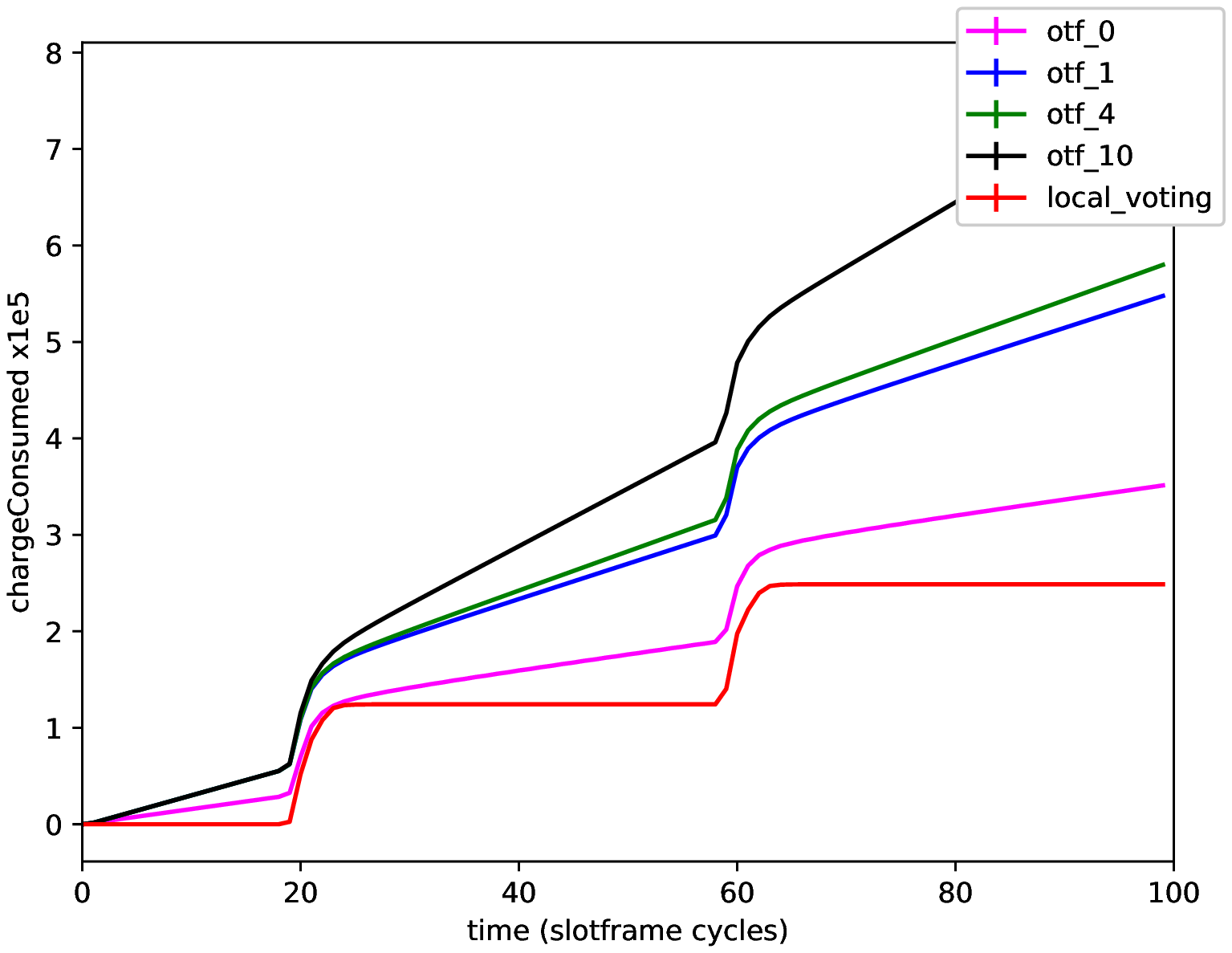}} 
	\caption{Energy consumption in \si{\micro \coulomb} over time for queue length of 100 packets when each node generates 5 packets per burst for (a) 1 parent, (b) 2 parents, and (c) 3 parents.}
	\label{energy10}
\end{figure*} 

The 6TiSCH simulator is an open-source discrete-event simulator written in Python by the members of the 6TiSCH WG \cite{simul}.
It implements the protocols: IEEE 802.15.4e-2012 TSCH \cite{MACstandard}, RPL \cite{rfc6550}, 6top \cite{wang-6tisch-6top-sublayer-04}, and OTF \cite{7273816}.
In addition to these protocols, we have added Local Voting \footnote{As an online addition to this article, the source code is available at \url{https://github.com/djvergad/local_voting_tsch}} as part of the work presented in this article. Since OTF has been already implemented in the 6TiSCH simulator \cite{simul}, we compare LV with OTF \cite{7273816}. We choose four threshold values for OTF (0, 1, 4, 10 cells) in order to provide thorough performance examination and comparison.
We work with the same simulation parameters as in \cite{7273816} which have been set according to RFC5673 \cite{rfc}. 
The simulation parameters are summarized in Table \ref{parameters}. The parameters are set according to an industrial environment scenario where traffic can be bursty. For instance, when detecting a leakage in an oil and gas system, the sensors transmit at a higher sample rate in order to minimize the time for detection of the leakage location, to calculate its magnitude, and to estimate the impact and the evolution of the leakage.
\shorten{
The network consists of $50$ nodes, randomly placed in an area of $2km \times 2km$. 
Every link is associated with a PDR value between 0.00 and 1.00. The PDR value per link is constant during a simulation run. Each node has at least three neighbors with links with PDR at least 50\%. A node is moved until this condition is satisfied.
The minimum acceptable RSSI value that allows for a packet reception is $-97dBm$, while the maximum number of MAC retries is set to 5.
The TSCH schedule contains 101 cells where each time slot has duration of $10ms$.
Each node generates data packets in bursts after $20s$ and $60s$ from the beginning of the simulation. We perform simulations where each node generates either 5 or 25 packets per burst. The queue length of all nodes is 100 packets.
The presented results are averaged over 500 simulation runs. A new topology is used for each run.} 

\begin{table}[!t]
	\renewcommand{\arraystretch}{1.3}
	\caption{Simulation Setup}
	\label{parameters}
	\centering
	\begin{tabular}{l  l}
		\hline
		\bfseries Parameter & \bfseries Value\\
		\hline
		Number of Nodes & 50\\
		Deployment area & square, $2km \times 2km$\\
		Deployment constraint & 3 neighbors with PDR>50\% \\
		Radio sensitivity &  $-97dBm$\\
		Max. MAC retries & 5\\
		Length of a slot frame & 101 cells\\
		Time slot duration & $10ms$ \\
		Number of channels & 16\\
		Burst timestamp& $20s$ and $60s$ \\
		Queue length & 100 packets\\
		Number of runs per sample & 500\\
		Number of cycles per run & 100\\
		6top housekeeping period & $1s$  \\
		OTF threshold & 0, 1, 4, 10 cells\\
		OTF housekeeping period & $1s$ \\
		RPL parents &  1, 2, and 3\\
		\hline
	\end{tabular}
\end{table}

Fig. \ref{reliability} shows the number of packets generated in the network that reach the root for queue length of 100 when each node generates 5 or 25 packets per burst. 
LV provides higher or similar level of reliability (a bigger portion of packets reach the root) than OTF for all threshold values. The number of packets that reach the root is significantly bigger with LV compared to OTF when the number of parents is 1. 
The time needed the packets to reach the root increases with decreasing the OTF threshold. As it is presented in Fig.~\ref{timeLast}(a), it takes longer time until the last packet reaches the root for OTF threshold equal to 0 compared to all other cases. 
LV performs always better in terms of both time for last packet to reach the root and end-to-end latency (Fig.~\ref{timeLast}(b)) compared to OTF for various values of the simulation parameters.
The end-to-end latency reduces for smaller buffer sizes and more parents for OTF while that is not always the case for LV. The results show that the latency reduces with increasing the number of parents for OTF, while the number of parents does not have a big impact on the latency for LV.
The total energy consumption of LV is also better than OTF for most of the scenarios (Fig.~\ref{timeLast}(c)).
The energy consumption over time is illustrated in Fig. \ref{energy10}.
LV consumes significantly less energy compared to OTF for all threshold values. The activity of LV is increased at $20s$ and $60s$, and hence the energy consumption goes higher at these timestamps. The energy consumption increases with the threshold for OTF. For instance, LV and OTF for threshold equal to 4 cells (queue length = 100, parent size = 3, 5 packets per node per burst) provide a similar level of reliability but both the end-to-end latency and energy consumption are two times higher with OTF compared to LV. The presented simulation results show that LV provides reliability close to or better than OTF while consuming less energy and providing lower end-to-end latency.

\section{Conclusions} \label{conc}
We proposed a new distributed link scheduling algorithm called \emph{Local Voting}. Local Voting allocates resources in the network by balancing the load between links in the network. In this way, it adapts the schedule to the network conditions in 6TiSCH networks and provides efficient resource allocation. Extensive simulation results show that in general the end-to-end latency is lower with Local Voting compared to OTF with different threshold values. Additionally, the number of packets that reach the root is higher and the energy consumption is lower with Local Voting compared to OTF. 

\bibliographystyle{plain}
\bibliography{refer}

\end{document}